\documentclass[epj]{webofc}
\usepackage[utf8]{inputenc}
\usepackage[varg]{txfonts}   
\usepackage{booktabs}
\usepackage{xcolor}
\definecolor{darkred}{rgb}{0.4,0.0,0.0}
\definecolor{darkgreen}{rgb}{0.0,0.4,0.0}
\definecolor{darkblue}{rgb}{0.0,0.0,0.4}
\usepackage[bookmarks,linktocpage,colorlinks,
    linkcolor = darkred,
    urlcolor  = darkblue,
    citecolor = darkgreen]{hyperref}
\usepackage{subfigure}
\wocname{EPJ Web of Conferences}
\woctitle{Lattice2017}
\begin{document}
%
\selectlanguage{english}
\title{%
Scientific and personal recollections of Roberto Petronzio
}
\author{%
\firstname{Giorgio} \lastname{Parisi} \thanks{Speaker, \email{giorgio.parisi@gmail.com} }
}
\institute{%
Dipartimento di Fisica,  Universit\`a di Roma {\it La Sapienza}, Piazzale A. Moro 2, I-00185, Rome, Italy
\and
INFN-Sezione di Roma 1, Piazzale A. Moro 2, I-00185, Rome, Italy
\and
Nanotec-CNR, UOS Rome, Piazzale A. Moro 2, I-00185, Rome, Italy
}
\abstract{%
  This paper aims to recall some of the main contributions of Roberto Petronzio to physics, with a particular regard to the period we have been working together. His seminal contributions cover an extremely wide range of topics: the foundation of the perturbative approach to QCD, various aspects of weak interaction theory, from basic questions (e.g. the mass of the Higgs) to lattice weak interaction, lattice QCD from the beginning to most recent computations.}
\maketitle
%
\section{Introduction}
Roberto Petronzio was born in 1949. He graduated with {Nicola Cabibbo} in 1972 in Rome University {\sl La Sapienza}, where he was a researcher up to 1979.  At that time he went to {CERN as staff member} where he remained from 1979 to 1986. 
 
 In 1987 he came back to Rome from CERN: he became a full professor of the newly founded University of {\sl Tor Vergata}. He accepted to take responsibilities in the direction of this university and for many years he served as vice-rector.
 
 From 2004 to 2011 he was {president of the INFN}, the Italian Institute of Nuclear Physics. This was a very demanding position:  he was forced to strongly reduce his personal scientific activity, although he succeeded from time to time to work on physics problems.  He was very successful in this difficult enterprise.  He managed to lead the institute through a very turbulent period, maintaining {independence from political power}, with the same calm and determination that he was using to safely lead a sailboat in a stormy sea. The Italian scientific community is very grateful to him for having devoted so much of his time to the collective interests.
 
 In 2011 he became president of the "Cabbibo Lab", a consortium that was supposed to construct a B-factory (the super-B project); this project was later canceled by the Italian government.
 
 Unfortunately, in 2014 he had a dramatic health accident and he died in July 2016.

 I will try to divide his works into different  categories that  partially overlap both chronologically  and  scientifically.
 \begin{itemize}
 \item {Perturbative QCD.}
 \item Weak interaction in the continuum.
 \item {Lattice QCD, the exploratory age.}
 \item Weak interactions on the lattice.
 \item {Lattice QCD, the mature age.}
 \end{itemize}

Roberto worked also on many other problems.  I will just mention a few:
\begin{itemize}
\item He was fascinated by the real space renormalization group in statistical mechanics \cite{Callaway:1984bq,Benzi:1988yd}. His contributions were quite varied: they ranged from first-order transitions \cite{Bernaschi:1989dm} to spin glasses \cite{parisi2001renormalization}.
\item He was very interested in biophysics: he obtained very nice results on the study of the folding of the Alzheimer peptide A$\beta_{16-22}$ \cite{rohrig2006stability}. This is a very important problem because this peptide is at the roots of the illness: this study was an important step forward, also for the new kind of computational techniques used.
\item He also worked on scattered problems of high-energy phenomenology. A few examples are radiative corrections in $e^+e^-$ annihilation \cite{Altarelli:1975za} or in neutrino scattering \cite{DeRujula:1979grv}, the effect of an hypothetic nearly massless particle on light propagation \cite{Maiani:1986md}, the design of a High Luminosity Tau/Charm Factory \cite{Biagini:2014dqa} \dots
\end{itemize}
\section{Pertubative QCD}
\subsection{The Roman period}
The first paper of Roberto was {\sl The nucleon as a bound state of three quarks and deep inelastic phenomena}. It appeared in August 1973 \cite{Altarelli:1973ff}.  It was based on the very nice idea of describing the quarks wave function inside the nucleon in the $p=\infty$ frame using information coming from internal symmetries like $\mbox{SU}(6)$. The results of this paper were later extended in \cite{Altarelli:1974is, Altarelli:1975qw} in order to get predictions for other processes like neutrino scattering and lepton production in proton-proton collisions.

It was a very interesting  paper for the following reasons: \begin{itemize}
\item Good models for the parton distribution were quite rare at that time. The paper describes the first realistic model valid not only for the quark structure functions, but for also for the {gluonic structure function}. The gluonic structure function will be later crucial for computing scaling violations via the process of gluon-fragmentation into quarks.
\item The model incorporates the knowledge that that time people had on symmetries, not only $\mbox{SU}(3)$, but also  $\mbox{SU}(6)_W$.
\item It stresses the importance of the  $p=\infty$ frame, that will play a very important role in understanding scaling violations in the framework of the extended parton model during subsequent years.
\end{itemize}

The paper assumed that the physical octet of Baryons is the combination of octets belonging to the following representations of $\mbox{SU}(6)_W$, the $56,\, l=0$ and the $70,\,l=1$. Let us consider  the case of a nucleon with spin component  $J_z=1/2$; this Baryon should be a linear combination of the following three states:
\begin{eqnarray}
 \label{e.all}
     |A\rangle&=& |8,\frac12,\frac12,0,0\rangle_{56}\\
 |B\rangle&=&-\frac{1}{\sqrt{3}}|8,\frac12,\frac12,1,0\rangle_{70}
 -\sqrt{ \frac23}|8,\frac12,-\frac12,1,1\rangle_{70}\\
 |C\rangle&=&\frac{1}{\sqrt{2}}|8,\frac32,\frac32,1,-1\rangle_{70}+
 \frac{1}{\sqrt{3}}|8,\frac32,\frac12,1,0\rangle_{70}+
 \frac{1}{\sqrt{6}}|8,\frac32,-\frac12,1,1\rangle_{70}
\end{eqnarray}
If only the 56 representation were present, one would obtain the bound
\begin{equation}
\frac32\ge\frac{F_2^{eN}}{F_2^{eP}}\ge\frac23 \,,
\end{equation}
that is violated by the experimental data, hence the need of introducing the mixing with the 70 representation. At the end it was possible to have a quite accurate description of the structure functions in terms of only a few parameters.

Our {first paper together} \cite{Altarelli:1974wm} was quite unfortunate: {\sl Is the 3104 MeV vector meson the $\psi_c$ or the $W_0$?} It was signed by {G. Altarelli, N. Cabibbo, R. Petronzio, L. Maiani, G. Parisi} (it was the only paper written by all these authors together).
The paper presented a nice phenomenological analysis: at the end of the paper, we concluded that the 3104 MeV vector meson was the weak interaction meson $W_0$, an answer that is factually wrong, in spite of the elegant arguments in the paper and of the quality of the authors.

\begin{figure}[thb]
  \centering
  \includegraphics[width=.5\columnwidth]{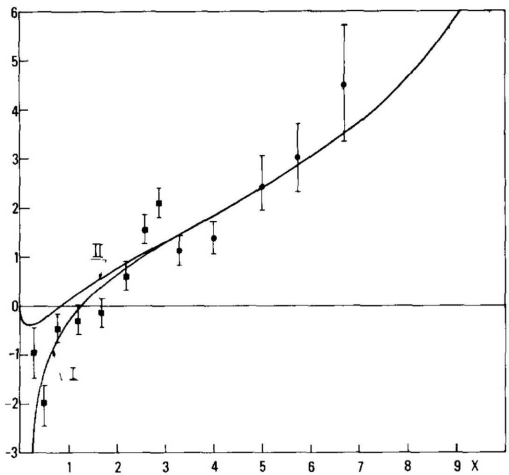}
 \caption{Curve I is our prediction for $d \ln F^p_x(x,q^2)/d \ln q^2$ compared with the experimental data. Curve II is obtained retaining only the octect operators in the operator expansion (taken from \cite{Parisi:1976fz}).
 }
 \label{PP76}
\end{figure}
 
 Our collaboration went on producing more interesting results. During the Roman period maybe our best paper together was {\sl On the breaking of Bjorken scaling} \cite{Parisi:1976fz}. This paper contains the first computation of scaling violations in QCD taking care of the presence of gluons, see fig. (\ref{PP76}). The paper was built on Roberto's great experience on parton wave functions inside the nucleon, especially on the gluonic contribution that was an essential component for having an agreement with the experimental results at small $x$: in this region gluon fragmentation into quarks is the dominant process.  It is remarkable that the computation was done 1976, before the AP (Altarelli Parisi) evolution equations \cite{Altarelli:1977zs}.
 
 Roberto Petronzio continued to work on the problem of scaling violations in deep inelastic scattering. Two years later he wrote with Nicola Cabibbo {\sl The Two-stage model of hadron structure: Parton distributions and their $Q^2$ dependence} \cite{Cabibbo:1978ez}, where a similar but more accurate analysis was done, now using the AP equations.

 \subsection{At CERN}   
 Roberto went to CERN in 1977 and spent 10 years there. Most of his works of the first years in CERN were on QCD and weak interactions. At that time the theoretical panorama on QCD was rapidly changing. The AP equations emancipated the study of partonic quarks from the need of considering the rather complex light cone expansion \cite{Brandt:1970kg} that played a crucial role in the initial period.  The AP equations study the evolution of the effective parton distribution inside a hadron.  This distribution is universal: it is the same in all the processes involving the same hadrons. However,  at that time it was realized that finite perturbative corrections proportional to the running coupling constant $\alpha(q^2)$ are present and they are process-dependent.
 
This global picture was easy to conjecture, but it was not easy to prove \cite{Veneziano:2017}. The proof finally came from a seminal work with deep theoretical consequences:  {\sl Relating hard QCD processes through the universality of mass singularities} \cite{Amati:1978wx}. It was the badly needed proof that the new approach was working: it was the key that opened the door to all the new developments.
  
 The point-like nature of QCD implied the existence of jets, of power tails in the transverse momentum distributions. However, at the time of that paper, the energy of the colliding particles was not high enough to see in a clear way the jets in the final states, also because the quark energy is partitioned between many hadrons via the process of jet fragmentation and quark recombination. On the other hand in the so-called Drell-Yan  process, i.e.
 \begin{equation}
p+p \to \mbox{hadrons}+\mu^++\mu^-\,,
\end{equation}
the transverse momentum of the $\mu^++\mu^-$ pair is equal to the transverse momentum of the quark antiquarks pair that produces the virtual photon: this process allows us to measure the transverse momentum spread or the quarks inside the proton.
 These considerations explain why hard scattering QCD contributions were firstly computed for the Drell-Yan process: the predictions were quite neat without having to discuss the process of quark fragmentation into hadrons. 
 
\begin{figure}[thb]
  \centering
  \includegraphics[width=.6\columnwidth]{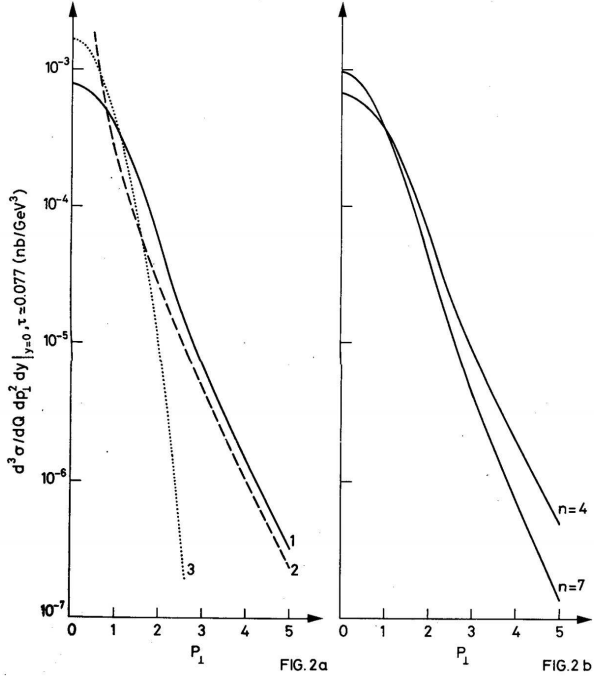}
 \caption{The differential cross section for the Drell-Yan process as function of $p_T$ (taken from \cite{Altarelli:1978pn}). Fig. 2a:  curve (1) is our prediction, curve (2) is the one loop contribution and curve (3) is the intrinsic contribution.  Fig. 2b shows how the results depend on the details of the gluon distribution, parametrized by $n$.
 }
 \label{APP78}
\end{figure}

 Two crucial seminal contributions were given by Roberto  in 1978 with the papers  
 {\sl Transverse momentum of muon pairs produced in hadronic collisions} \cite{Altarelli:1977kt} and {\sl Transverse momentum in Drell-Yan processes} \cite{Altarelli:1978pn}. A careful job was done in studying the increase of the average transverse momentum squared ($p_T^2$)  as a function of $Q^2$ and of the various physical parameters: some of the results are shown in fig. (\ref{APP78}).
 A problem that we had to face was the separation of the two contributions: the one coming from intrinsic spread of the quark wave function inside the nucleon and the one coming from hard processes.

 Roberto was very interested in the resummations of leading logs in special processes, a problem that was studied in the case of QED, but not for QCD. The first paper on this subject is {\sl Heavy flavor multiplicities at very high energies} \cite{Furmanski:1979jx}. New techniques had to be invented in order to circumvent new difficulties.
 The authors found the surprising result that the multiplicities increase faster than any power of the logarithm of the energy scale, i.e. they derived this behavior
\begin{equation}
\langle n\rangle \propto \exp\left( \sqrt{\alpha_G \log(Q^2/Q_0^2)}\right)\,.
\end{equation}
This result was found quite puzzling by the author themselves, and this reaction was natural: at that time a simple logarithmic increase of multiplicities was supposed to be experimentally established. Nowadays  we know experimentally that the multiplicities increase much faster than a logarithmic of the energy and the results are much less puzzling.
 
A paper that had a long influence was
 {\sl Small transverse momentum distributions in hard processes} by Roberto and myself \cite{Parisi:1979se}.  We wanted to find the {\sl small} transverse momentum behavior of the distribution of hard produced muon pairs as an effect of multiple gluon production. In the computation done in \cite{Altarelli:1977kt,Altarelli:1978pn} an intrinsic momentum distribution was needed to avoid the singularities at $p_T=0$ that are present in the first order in perturbation theory. From the physical viewpoint it was clear that multiple gluon production should produce a regularization effect at small momentum, however, the detailed consequences of this phenomenon were not clear.
 
Many ingredients entered in the cocktail  \cite{Parisi:1979se}. 
\begin{itemize}
\item The leading logs approximation for multiple soft gluon bremsstrahlung.
\item The exponential damping of the elastic form factors.
\item The different behavior of the cross sections in momentum and in impact parameter space.
\end{itemize}
 One of the conclusions of that paper (that I still find surprising) is that the peak at $p_T=0$ flattens with a width proportional to $\left(Q^2\right)^{\gamma}$ with $\gamma=\frac{16}{25}\ln(66/41)\approx 0.305$. The presence of a simple non-integer power of $Q^2$ is quite astonishing in a world dominated by logarithmic corrections.
 
 Another paper that had a long and larger influence was
 {\sl Singlet parton densities beyond leading order} \cite{Furmanski:1980cm}.
 This was the {\sl manifesto} for next to the leading order computation in QCD. The difficulties tacked in this paper were not only related to doing the detailed computations, that were highly nontrivial; the main point was to prove for the first time that those computations were possible.
 The technical tool that they used was based on the explicit study of the factorization properties of mass singularities that was already present in  \cite{Amati:1978wx}. The conclusions of the paper were that in this way {\sl "within our scheme the predictions for a particular process are obtained by convoluting a universal parton density with a short-distance cross-section specific to the process."} It was the triumph for the marriage of the parton model with QCD.

 These results were extended  to the non-singlet cases in the papers 
 {\sl Evolution of parton densities beyond leading order: The non-singlet case} \cite{Curci:1980uw}  and 
 {\sl Lepton-hadron processes beyond leading order in quantum chromodynamics} \cite{Furmanski:1981cw}.
  \begin{figure}[thb]
  \centering
  \includegraphics[width=.55\columnwidth]{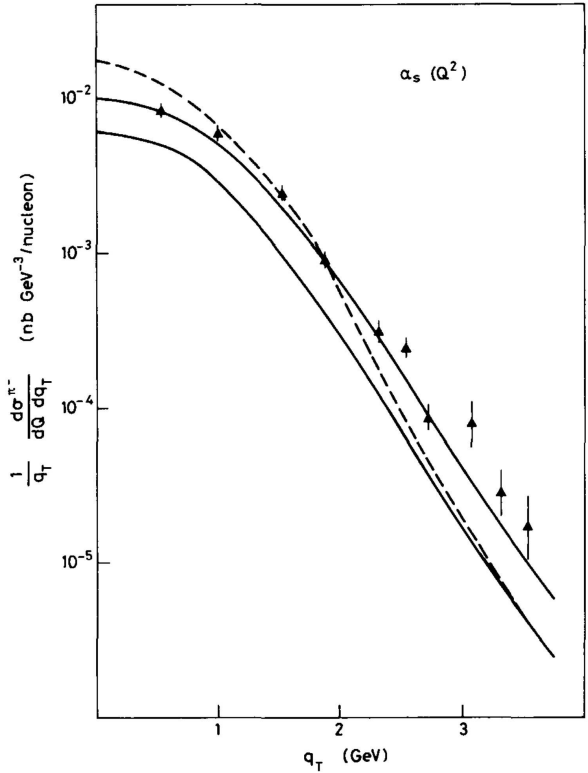}
 \caption{ Experimental data for the Drell-Yan process in $\pi^-$ nucleon collisions (i.e. $1/(q_T)d\sigma^{\pi^-N}/(dQ\,dq_T)$) versus $q_T$ at $s=282\, \mbox{GeV}^2$, $Q^2=52.2\,\mbox{GeV}^2$,  $Q^2$ being the invariant mass of the muons pair.   The data are compared with the theoretical predictions: the upper full line represents the next-to-leading estimate while the lowest full line represents the lower order estimate (taken from \cite{Ellis:1981hk}).
 }
 \label{EMP}
\end{figure}
 The  techniques introduced in these papers allowed the computation of next to the leading order results in other processes, as it was done in  the paper by 
 {\sl Lepton pair production at large transverse momentum in second order QCD} \cite{Ellis:1981hk}. This is a very remarkable paper, because it contains the first evaluation of next to the leading order effects in QCD for the $p_T$ distribution in the Drell-Yan process. The computation was quite involved because the authors had to compute the $\alpha_s^2$ corrections to a process of order $\alpha_s$. You can see from fig.  (\ref{EMP}) the importance of adding next to the leading effects in order to reach a good agreement with the experimental data.

 Roberto continued to gave seminal contributions to QCD with {\sl Power corrections to the parton model in QCD} \cite{Ellis:1982wd} and
 {\sl Unravelling higher twists} \cite{Ellis:1982cd}.
 A paper that was quite ahead of its time was
 {\sl Momentum distribution of $J/\Psi$ in the presence of a quark-gluon plasma} \cite{Karsch:1987uk}. The subject of the paper is quite different from the previous ones.  In heavy nuclei collisions at very high energies we could have the formation of a new phase of matter, i.e. {\sl quark-gluon plasma}: there were many theoretical arguments that pointed in that direction. However, it was not clear which was a good experimental signature for this phenomenon. In this paper, the authors presented for the first time their very interesting suggestion that the momentum distribution of the produced $J/\Psi$ particles should be strongly affected by the phase transition to this new state of matter.
 
 \section{Weak interactions and supersimmetry}
 
 Roberto was always very interested in weak interactions. There were two remarkable papers written at the end of the seventies:
  {\it Bounds on the Number and Masses of Quarks and Leptons} \cite{Maiani:1977cg} by Maiani, myself and Roberto  and
  {\it Bounds on the Fermions and Higgs Boson masses in grand unified theories} \cite{Cabibbo:1979ay} by Cabibbo, Maiani, myself and Roberto.  Luciano Maiani remembers \cite{Maiani:2017} that we three had many discussions (some of them via handwritten mail) on the subject concerning the first paper, but we were far from concluding. Once he went to CERN to discuss the matter with Roberto: Roberto put all the results we had on the table. He argued that the physical picture was quite clear if we combined all the information we had: after a few hours of discussions the skeleton of the draft of the paper was written.
  
\begin{figure}[thb]
  \centering
\includegraphics[width=0.5\textwidth]{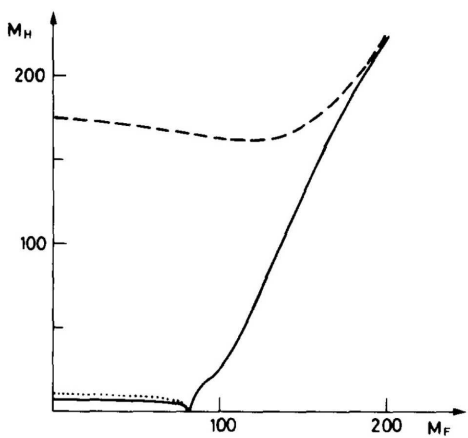}
\caption{Bounds on the mass of the Higgs Boson ($m_H$) as a function of the top quark mass ($M_F$) in the case of three generations for $\sin^2(\theta_W)=0.2$. The dashed line and the full line represent the upper and the lower bound respectively and the dotted line is the prediction of the massless theory. The curves end in correspondence to the upper bound on $M_F$. (Taken from \cite{Cabibbo:1979ay}.)}
\label{CMPP}
 \end{figure}

  The conclusion of the two papers was "In the framework of grand unifying theories, the requirement that no interaction becomes strong and no vacuum instability develops up to the unification energy is shown to imply upper bounds to the Fermion masses as well as upper and lower bounds to the Higgs boson mass." In the same way, a bound on the value of the top mass was obtained. The original bounds are shown in fig. (\ref{CMPP}) as a function of the top mass: the top mass was not known at that time and in the paper we derived an upper bound of about 220 GeV.  Using the actual value of the top mass (i.e. 172 GeV), the bounds on the Higgs mass was quite sharp.
  
 These bounds were quite good: they were based on the leading order for the evolution of the coupling constants: more precise and accurate bounds have been found by refining the computation by including high order terms. It is a very interesting fact, whose significance is not clear, that the experimental value of the Higgs meson is quite near to the lower bound.
 
 Many years later Roberto came back to the study of weak interactions in the continuum when he started to be strongly interested in the fascinating problem of the supersymmetric extensions of the standard model. 
 
 A remarkable paper is
 {\it Flavour changing top decays in supersymmetric extensions of the standard model} \cite{deDivitiis:1997sh}. In this paper, it was noticed the flavor changing top decays $top \to charm +Z_0$, $top\to charm +g$, $top\to charm +\gamma$,  are predicted with invisible rates within the standard model and may represent a window on new physics. These processes have been considered in supersymmetric extensions of the standard model: the authors showed that observable rates can be obtained only if the SUSY breaking is non-universal and flavor dependent.

Another very interesting paper of a few years later  is  {\it Probing new physics through $\mu-e$ universality in $K\to l+\nu$} \cite{Masiero:2005wr}  (2006). In this paper, it was shown that supersymmetric (SUSY) extensions of the standard model can exhibit   $\mu-e$
 non-universal contributions. They are quite effective in constraining relevant regions of SUSY models with lepton flavor violating currents. This work was done when Roberto was president of the INFN and Roberto had to do some tricks (e.g. hiding with Masiero in a utility room) to isolate himself from the tasks related to his office \cite{Masiero:2017}.

 \section{Lattice QCD, the exploratory age.}
 
 At the beginning of the eighties, the central interests of Roberto already started to move toward lattice theories and lattice QCD. 
 The subject was completely new and there was the need to understand which were the possible artifacts of lattice computations.  The simplest case was the two-dimensional O(3) spin model. The physics of the model was very clear (a ferromagnetic transition that was avoided as an effect of the impossibility of having a Goldstone mode in two dimensions). Moreover, the theory was asymptotically free (like QCD) and topological effects, like instantons, were present also in this case.  The simplicity of the theory allowed many detailed computations.

 \begin{figure}[thb]
  \centering
  \includegraphics[width=.5\columnwidth]{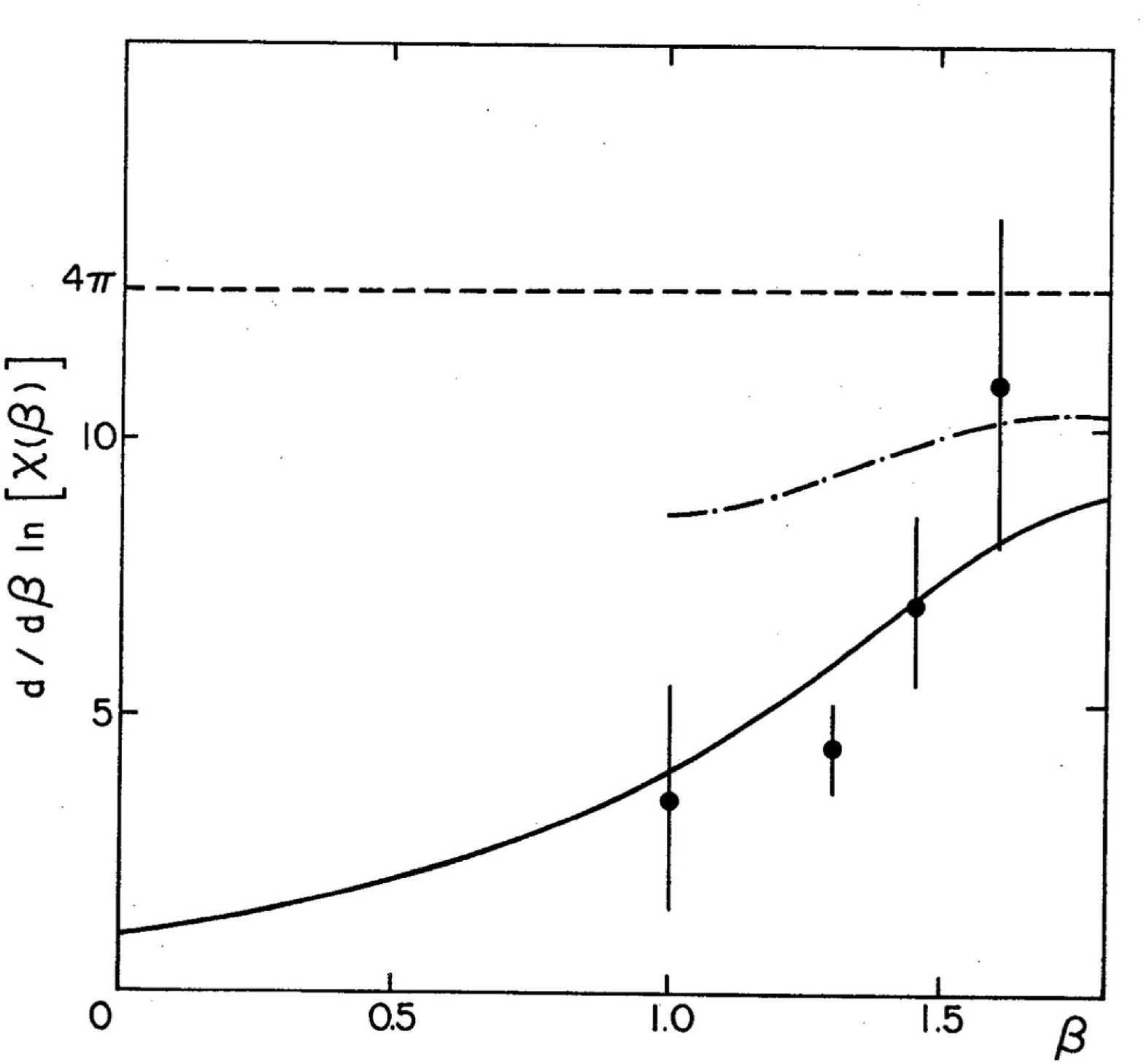}
 \caption{ The quantity $d\log(\chi(\beta))/d\beta$ as function of $\beta$. The dashed line is the asymptotic value at $\beta=\infty$; the dashed dotted curve takes care of the preasymptotic corrections coming from the next to the leading order (taken from \cite{Martinelli:1980tb}).
 }
 \label{MPP}
\end{figure}

 Roberto, Martinelli and I started to perform Montecarlo simulations for this theory. In our first paper, {\sl Monte Carlo simulations for the two-dimensional $O(3)$ nonlinear sigma model} \cite{Martinelli:1980tb}, we tried to study for the first time \footnote{In that years Montecarlo for lattice theories was such a new approach that most of the computations were done for the first time.} the behavior of the magnetic susceptibility $\chi(\beta)$ at high $\beta$ (low temperature). 
 We knew from analytic computation that for large $\beta$ 
 \begin{equation}
\chi(\beta) \propto \beta^{-4}\exp(4\pi \beta) \qquad \mbox{hence} \qquad \lim_{\beta\to\infty}\frac{d\log(\chi(\beta))}{ d\beta}=4\pi
\end{equation}

 We wanted to understand how fast the limit was reached and we were not happy because the approach was quite slow. We used lattices with $L^2$ points with $L$ in the range from 30 to 80 and the results are shown in fig. (\ref{MPP}). It was clear that we needed a much larger lattice in order to be near to the asymptotic limit. The conclusions were quite scary: if the same phenomenon was present for QCD, the whole field would be have been destroyed for a few decades: four-dimensional lattices with $L=80$ are at the boundary of present-day technology.
 
 In order to decrease lattice effects for this model, we found  which was the form of the improved lattice action  where $O(a^2)$ corrections were absent \footnote{As usual, $a$ is the lattice spacing.}:  this was done in {\sl Improving the lattice action near the continuum limit} \cite{Martinelli:1982db}. This computation was done taking care also of one loop corrections that had to be evaluated for the lattice theory. In some sense, we computed the difference between the one loop results in the continuum and one loop results in the lattice: at the end of the day, we added counter-terms in order to compensate for the difference of these two computations. This was the first of a huge family of improved actions that have been widely used in QCD and in weak interactions on the lattice.
  
 Another remarkable paper of that time was {\sl Topological charge on the lattice: The $\mbox{O(3)}$ case} \cite{Martinelli:1981pb}.  In this paper, the authors presented the first definition {the topological charge} for the two-dimensional O(3) spin model, that had the property of being insensitive to small instantons. The instanton density was computed and it was compared with the analytic results.

 However, most of the fun was with lattice QCD.
 It was a new world that we started to explore with excitement. All the low-energy strong interaction parameters were computable. This was a complete change from the previous situation where only phenomenological arguments could be used, mostly in hand-waving arguments. Of course we knew that the measurements were affected by strong systematic effects (we started our computation with a $5^3\times 10$ lattice), however, it was rather surprising to see that all quantities, one after the other, were in qualitative agreement with the experimental data.
 
 There are so many papers in that period that I will just briefly recall them. The collaboration was floating and the author list often changed.
 \begin{itemize}
 \item We started our collaboration with the computation of the basic properties of hadrons in the quenched approximation in {\sl Hadron spectroscopy in lattice QCD} \cite{Fucito:1982ip}, where the statistic and systematic errors were strongly reduced with respect to the previous papers.
 \item We computed the {\sl  proton and neutron magnetic moments in lattice QCD} \cite{Martinelli:1982cb} by measuring the mass splitting in presence of a magnetic field: in this case we found for the gyromagnetic factor of the proton the value $g_P=3.0\pm 0.6$ versus an experimental value of 2.79 and for the ratio of the gyromagnetic factors of the proton and of the neutron $g_P/g_N=-1.60\pm 0.15$ versus an experimental value of $-1.46$.
 \item We computed  the strange hadron masses \cite{Martinelli:1982eu}, in particular the {\sl $\Lambda-\Sigma_0$ splitting}. Here the result was not too satisfactory: the sign was the correct one, but its absolute value was quite small. We argued that this was an example of a general phenomenon: all the mass splitting due to spin-spin interactions were quite small. We obtained a reasonable value to the ratio
 \begin{equation}
\frac{m_\Sigma-m_\Lambda}{m_\Delta - m_P}=0.18\pm 0.09\,,
\end{equation}
to be compared to the experimental value of 0.26.
\item  In {\sl Boundary effects and hadron masses in lattice QCD} \cite{Martinelli:1982bm} we identified a relevant contribution to the large fluctuations of hadron masses present in lattice calculations with periodic boundary conditions.  This contribution is due to unphysical quark paths which are absent in the infinite volume limit. We showed that these contributions can be eliminated by averaging over possible rotations of the boundary links by the elements of the $Z(3)$ subgroup. In this way, the  "effective" volume for these paths is triplicated.
 \end{itemize}

A very remarkable paper was {\sl Hadron spectrum in quenched QCD on a $10^3\times 20$ lattice} \cite{Lipps:1983pi} by Lipps, Martinelli, Petronzio and Rapuano.  It was real progress respect to the previous analysis on smaller lattices ($5^3$, $6^3$, $8^3$) and allowed us for the first time to investigate the systematic effect due to non-zero lattice sides.  A subsequent paper was {\sl Kogut-Susskind and Wilson fermions in the quenched approximation: A Monte Carlo simulation} \cite{Billoire:1984jm} where the authors presented a systematic comparison of the results for both the Kogut-Susskind and Wilson Fermions.

Roberto was also interested to analyze the behavior of QCD without Fermions. Here the most relevant observable (beyond the glueball mass) is the string tension. However, his precise determination was quite difficult due to large statistical errors. In spite of these difficulties in \cite{Parisi:1983hm} we computed the string tension with good accuracy. The computation was possible due to a clever trick for noise reduction (i.e. multihit) that we introduced in that paper and that became a standard tool. The gain induced by the trick was a decrease of a factor 10 in the statistical error, corresponding to a gain of a factor 100 in time. Roberto continued to work on pure gauge QCD. He wrote a very nice paper {\sl Gluon thermodynamics near the continuum limit} \cite{Karsch:1983ag} on the quark liberation phase transition (that correspond to the formation a quark-gluon plasma), a problem that we have already seen he analyzed in a subsequent paper with the same author \cite{Karsch:1987uk}.

Roberto was also among the proponents of the first  APE project. He contributed to the first two papers presenting the design of  the first APE computer: {\sl The APE project: a computer for lattice QCD} \cite{Bacilieri:1984ai} and {\sl The APE project: a gigaflop parallel processor for lattice calculations} \cite{Bacilieri:1985pa}. Unfortunately, the collaboration with Roberto inside the APE project could not continue due to logistic problems. During the construction of the machine the work was concentrated in Bologna (memory card), Pisa (controller and local network), Rome (floating point unit and software), but not in CERN. He was strongly involved in using the subsequent APE machines, but this part of the story will be discussed later.

\section{Weak and electromagnetic interactions on the lattice}

In 1983 a new investigation subject in lattice gauge theories was open with the paper {\sl Weak interactions on the lattice} \cite{Cabibbo:1983xa}: the authors showed that lattice QCD can be used to evaluate the matrix elements of four-Fermion operators which are relevant for weak decays. This was the starting point of so many computations of weak matrix elements that are very important in the testing of the standard model and in the eventual discovery of new physics. 

The conclusion was that lattice QCD can be used to evaluate the matrix elements of four-Fermion operators which are relevant for weak decays.
Indeed in the paper, we find many results of the kind:
\begin{equation}
\frac{\langle \pi^+|\bar (u \gamma^\mu_L u)(\bar s \gamma^L_\mu d)|K^+\rangle}
{m^2}= 4a^2 (3.7\pm 0.3)10^{-3}  \qquad [2.4\ 10^{-3}]\,,
\end{equation}
where the value in brackets is given on the basis of a theoretical approach based on vacuum saturation of some QCD sum rules. The authors were happy because there was a general agreement between their results and those coming from vacuum saturation of sum rules that was the most established approach at that times. The method was giving very promising results: it was the starting point of many many investigations.

After this seminal work, Roberto returned to the study of weak interactions on the lattice only in the nineties. In this period, among his first papers on the subject, we find: {\it Dynamical flavor dependence of static heavy meson decay constants on the lattice} \cite{deDivitiis:1996yw}. This is the first paper of a series devoted to  the computation of heavy meson decay: {\it Heavy quark masses in the continuum limit of quenched lattice QCD} \cite{deDivitiis:2003iy} and
{\it Quenched lattice calculation of the $B\to D\, l \,\nu$ decay rate} \cite{deDivitiis:2008df}. His strong interest in heavy meson decay was clearly triggered by the need of firm theoretical predictions in order to put strong constraints on possible violations of the standard model predictions.

Roberto has been always looking for very reliable predictions: he has always been very concerned about systematic errors. If the predictions were affected by uncontrolled systematic errors they were useless to discover new physics. For these reasons, he wanted to close all possible loopholes. In this regard a very important paper is  {\it  
Nonperturbative renormalization constants on the lattice from flavor non-singlet Ward identities} \cite{deDivitiis:1997ka}, where the authors obtain absolute predictions for the normalization of the current on the lattice.

In more recent years, when times were mature, he started to be interested not only in weak interactions but also in full electromagnetic interactions looking for the effects of this interaction on violations of isospin symmetry. In a first paper,  
 {\it Isospin breaking effects due to the up-down mass difference in Lattice QCD} \cite{deDivitiis:2011eh} the authors consider mainly the effects of the mass differences of the quarks. 
 
 As it is well known the difference in masses of the quarks is not sufficient to explain all the isospin violations (e.g. the mass difference between the $\pi^+$ and the $\pi^0$). The effect of the real electromagnetic field has to be taken into account and this was done in a later paper, {\it Leading isospin breaking effects on the lattice} \cite{deDivitiis:2013xla}. 
The authors presented a method to evaluate on the lattice the leading isospin breaking effects due to {\it both} {the small mass difference} between the up and down quarks and the {QED interaction}. 
They treated the dynamical quarks as electrically neutral particles ({\it electroquenched} approximation) and computed the charged and neutral pion mass splitting neglecting only a disconnected diagram. The final results were in very good agreement with the experimental data also for this problem that was rarely investigated before.

 \section{Lattice QCD, the mature age.}
 
Roberto started again to work on large-scale simulations of pure QCD in the nineties. The situation was changed from the time of the first works in the field of ten years before. Now the field was mature. Exploratory works were already done: there was the need of controlling well the sources of possible systematic errors. Moreover given the panoply of different forms of the lattice action, it was crucial to be sure that they gave asymptotically consistent results and that there were no lattice artifacts we were not aware of. Indeed the motivations were not so different from those of the paper \cite{deDivitiis:1997ka} that we have discussed above.

This motivation lead these two papers: {\it Non-perturbative determination of the running coupling constant in quenched SU (2)} (1993) \cite{deDivitiis:1993hj} and 
{\it Universality and the approach to the continuum limit in lattice gauge theory} \cite{deDivitiis:1994yz}. 
The results of this accurate and innovative analysis were that using a finite-size renormalization group technique it was possible to calculate the running coupling constant for quenched SU(2) with a few percent error over a range of energy varying by a factor thirty. They used a definition based on the ratio of correlations of Polyakov loops with twisted boundary conditions. At the end, the authors found that the extrapolation to the continuum limit was governed by corrections due to lattice artifacts which appear to be rather smooth and proportional to the square of the lattice spacing. 

\begin{figure}[thb]
  \centering
 \includegraphics[width=0.5\textwidth]{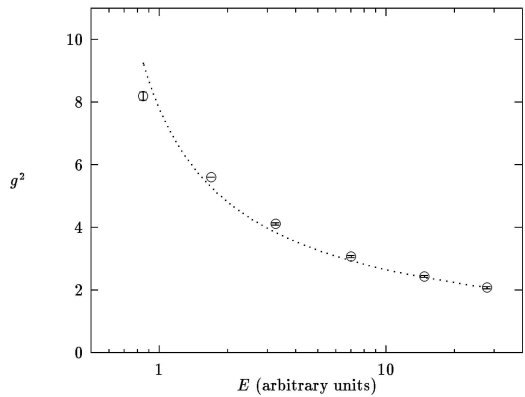}
 \caption{The running coupling constant as function of the energy: the dashed line is the result from the two loop perturbative computation (taken from \cite{deDivitiis:1993hj}).
 }
 \label{running}
\end{figure}

If we compare fig. (\ref{running}) with fig. (\ref{MPP}), that was done for a much simpler model, we have a vivid graphic account of the progress that has been done in a dozen of years.

Roberto continued to work on lattice QCD, in parallel with his works on weak interactions. A very interesting paper of 2004  is 
{\it On the discretization of physical momenta in lattice QCD} \cite{deDivitiis:2004kq} . As it is well known, the smallest physical momentum in a box of side $L$ is $2\pi/L$. The factor $2 \pi$ (not a very small number) lead to a large value of the minimum non-zero momentum. This a great nuisance when we are interested to compute the momentum dependence of some observable. 

This difficulty has been partially removed in this paper, where the authors showed that the limitation represented by the finite volume momentum quantization rule can be overcome by using different boundary conditions for different Fermion species. The very interesting conclusion was that the method proposed can be applied to study all the quantities of phenomenological interest that would benefit from the introduction of continuous physical momenta like, for example, weak matrix elements.

In spite of the heavy work duty as president of the INFN,  he continued to work on lattice QCD. A very remarkable paper of 2007 is  {\it QCD with light Wilson quarks on fine lattices: first experiences and physics results} \cite{DelDebbio:2006cn}. In this paper the universality of the continuum limit and the applicability of renormalized perturbation theory are tested in the SU(2) lattice gauge theory by computing {two different non-perturbatively defined running couplings over a large range of energies \cite{Lusher:2017}.
The lattice data (which were generated on the powerful APE computers at Rome II and DESY) are extrapolated to the continuum limit by simulating sequences of lattices with decreasing spacings. The results confirmed the expected universality at all energies to a precision of a few percent. The author found, however, that perturbation theory must be used with care when matching different renormalized couplings at high energies.

\section{Conclusions}
Roberto was a very talented physicist with a very strong physical intuition; he was also a dedicated hard worker (he published about 200 papers). He wanted to work on any problem that he found interesting. 
He did not give too much weight to the possible fame obtained by solving the problem, but he was driven by his immense curiosity and by the joy that one has in improving his understanding. It was the same attitude that led our common mentor, Nicola Cabibbo, to repeated often 
{\it Why should we work on this problem if we do not have fun?}

 Roberto was a very charismatic person and a great leader. He was also a highly appreciated science manager (often the two things do not go together).
 He had an exceptionally deep understanding of many fields of physics ranging from high energy to statistical mechanics.  Roberto was a great teacher: he was able to communicate his knowledge with great enthusiasm to young students and collaborators on which he had a lasting influence \cite{Sagnotti:2017}.

The quality of human relations with all the people he was in contact was very important for him: he paid lot of attention to be kind with all the people he met and not to delude their expectations (not an easy job when you are at the head of research institute where a few thousand people work). He had a sincere interest in other people difficulties and he usually did all he could do to help them, quite often with success. 

Everybody who met him will remember his smile, his positive attitude toward life. We will dearly miss him.

\clearpage
\bibliography{Lattice2017_indicoID_PARISI}

\end{document}